\documentclass[11pt]{article}
\usepackage{cospar}
\usepackage{url}

\usepackage{graphicx}
\usepackage{graphpap}
\usepackage{epsf}

\def\etal{et al. }
\def\nifsx{${}^{56}$Ni}
\def\msun{M_\odot}

\def\FIG #1 #2 [#3] #4\par{%
  \begin{figure}[!h] \begin{center}%
    \includegraphics*[#3]{#2}%
    \vskip -9mm
    \caption{\label{#1}#4}
  \end{center}\end{figure}%
}

\def\FIGG #1 #2 #3 [#4] #5\par{%
  \begin{figure}[!h]
    \vskip -3mm
    \includegraphics*[#4]{#2}
    \hfill
    \includegraphics*[#4]{#3}
    \vskip -10mm
    \caption{\label{#1}#5}
    \vskip -5mm
  \end{figure}
}

\def\rfig#1{Fig.\ref{#1}}

\renewcommand{\t}{\,\mbox}

\title{\uppercase{X-ray emission of young SN~Ia remnants \\
       as a probe for an explosion model}}

\author{D.I.~Kosenko\address{Sternberg Astronomical Institute, 119992,
Moscow, Russia}, E.I.~Sorokina$^{1}$,
S.I.~Blinnikov$^{1}$\address{Institute for Experimental and
Theoretical Physics, 117218, Moscow, Russia},
P.~Lundqvist\address{Stockholm Observatory, Albanova, Stockholm,
Sweden}}

\begin{document}

\maketitle

\begin{abstract}
We present results of hydrodynamical simulations of young
supernova remnants. To model the ejecta, we  use several models
(discussed in literature) of type Ia supernova explosions with
different abundances. Our hydro models are one-dimensional and spherically
symmetrical, but they take into account ionization kinetics with
all important processes. We include detailed calculations for the
X-ray emission, allowing for time-dependent ionization and
recombination. In particular, we compare the computed X-ray
spectra with recent XMM-Newton observations of the Tycho SN
remnant. Our goal is to find the most viable thermonuclear SN
model that gives good fits to both these X-ray observations
and typical SN~Ia light curves.
\end{abstract}

\section*{Introduction}

Recent observations with XMM-Newton telescope with high spectral
and space resolution, give us an opportunity to get clearer
understanding of nature and origin of X-ray sources. For example,
with new data on supernova remnants we can find out more about
the supernova explosion itself. So far, there exist many explosion models
proposed by theorists for different types of supernovae,
but still there are no definite criteria to decide which
of the models are realized in nature. During the first
months after an explosion one can examine a theoretical model by
calculating bolometric and monochromatic light curves and spectra.
Later on, gas in the ejecta cools down and becomes almost
unobservable. The next opportunity to analyze the ejecta is
on the stage of a young supernova remnant (SNR), when noticeable amount of
circumstellar gas is swept up. At this moment a reverse shock
forms, goes inwards the ejecta and illuminates it once again.

We will focus only on Type Ia Supernovae (SN~Ia) and compare the
results for four theoretical models of explosion. We are eager to find
a model which would be in good agreement with observations of
SN~Ia both at the epoch of maximum light and on the young remnant stages.

In this work, we calculate hydrodynamical evolution and X-ray
emission of a supernova remnant at the age of 430 years, which
corresponds to the age of Tycho SNR, and compare the results with
spectra and images of this remnant obtained by XMM-Newton space
telescope (Decourchelle et al. 2001). We examine the same models
that where applied to our broad-band light curve modelling (Blinnikov,
Sorokina~2003), so we
will be able to judge on the correctness of these models from two
points of view.

Quite a while ago, it was concluded that Tycho's supernova was of
Type~I (Baade 1945).
After that time there were many other papers which suggested other types
for Tycho.
But the morphology of the remnant, its shell-like structure, says that most
probably it is still the remnant of Type~Ia supernova.

The basic parameters of the Tycho remnant are the following: the age
is 430 years, the angular size is $\sim 8'$,
thus the radius is $\sim 2 - 3$ pc (depending on adopted distance).
It has almost spherical shape, so we do not need any additional assumptions
to model it with a 1D code.

Tycho's remnant was observed for many years in radio wavelength and optics
with terrestrial instruments (e.g., van den Bergh~1971,
van den Bergh et al.~1973, Duin, Strom~1975, Dickel et al.~1982) and in X-rays
with different space telescopes (e.g., Davison et al.~1976,
Seward et al.~1983, Tsunemi et al.~1986, Hwang, Gotthelf~1997,
Decourchelle et al.~2001).
Its images and spectra were analyzed in many papers (e.g., Chevalier,
Raymond~1978, Hamilton et al.~1986, Itoh et al.~1988, Brinkmann et al.~1989,
Vancura et al.~1995, Hwang et al.~1998).
The instruments become better every decade.
The most recent observations with XMM-Newton provide us with data of
excellent spatial resolution, which allows us to examine not only crude
models to reproduce the emission of the remnant, but also to change between
models which differ not very strongly.

\section*{Models}
To simulate supernova ejecta in our
calculations of evolution and X-ray emission of remnant we have chosen
four SN~Ia explosion models. Three of them are Chandrasekhar-mass models and
one is a sub-Chandrasekhar one:
\begin{itemize}
\setlength{\parskip}{0pt} \setlength{\itemsep}{0pt}
\item the classical deflagration model W7 (Nomoto et al. 1984).
$M_{WD}=1.38\msun$, $E_0=1.2\times10^{51}$~ergs, M(\nifsx) $= 0.6
\msun$.

\item the deflagration model by Reinecke et al. (2002)
 (hereafter MR0).
$M_{WD}=1.38\msun$,
$E_0=4.6\times10^{50}$~ergs, M(\nifsx) $ = 0.43 \msun$.

\item the delayed detonation model DD4 (Woosley \& Weaver 1994b).
$M_{WD}=1.38\msun$, $E_0=1.2\times10^{51}$~ergs, M(\nifsx) $ =
0.6 \msun$.

\item the sub-Chandrasekhar-mass detonation model with
low \nifsx\ production  (hereafter, WD065; Ruiz-Lapuente \etal
1993). $M_{WD}=0.65 M_{\odot}$, $E_0 = 5.6\times 10^{50}$~ergs, M(\nifsx) $
=0.5 \msun$.

\end{itemize}

W7, DD4, and WD065 are well known SN~Ia models. The last one is
the best for modelling light curves and late spectra of very
sub-luminous SNe~Ia, like 1991bg. Two others are usually
considered as the best models for typical SN~Ia.

The recent MR0 model is originally 3D, but we have averaged it over the
whole $4\pi$, since our hydro code is 1D. MR0 is almost a ``first
principle'' model, and uses much less free parameters for flame
modelling than all previous calculations. It is less energetic,
contains less \nifsx, but the latter is located in the outermost layers
of ejecta due to 3D Rayleigh--Taylor instability. The combination of all these
factors leads to the situation when the outermost \nifsx\ layer
has similar velocities for all Chandrasekhar-mass models.

From our light curve modelling (Blinnikov, Sorokina~2003) we have found
that MR0 model is the best for UBVI bands, though the bolometric light curve
is more similar to observations for W7.
We discussed there that a bit more energetic model, and equally mixed as MR0,
would be the best from the light curve modelling point of view.

We surround ejecta by a motionless gas of constant density ($5
\times 10^{-24} \mbox{ g/cm}^3 $), constant temperature
($10^4$~K), and with Solar abundances. We start calculations of the remnant
evolution
at the age of ejecta from several years to several decades.

\section*{Method}

To model the hydrodynamical evolution of the supernova remnant
we use hydro code from the package
STELLA (Blinnikov et al. 2000).
To find out the X-ray spectra we use
the code written by P.L. for calculations of collisional ionization of a
stationary plasma, which takes into account basic processes:
ionization by electron impact, autoionization, photorecombination,
dielectronic recombination, charge transfer. It calculates
bremsstrahlung (free-free), free-bound, bound-bound, two-photon
emission. This code has been elaborated by S.B. and
E.S. into a time-dependent variant ($\dot n_i =
f(T(t),n(t),Z,...)$ for all ionic species). Input parameters are current
temperature, density, abundance.

 The models were calculated as
follows. We simulate an SN expansion into
CSM, assuming that the remnant is transparent, i.e. radiation does
not affect remnant's dynamics, so we can calculate hydrodynamics
and radiation separately. During the calculation we record history
of temperature and density variations for each mesh zone and then
we use this data to calculate time-dependent ionization. The
resulting ionization stages for all elements at the age of Tycho
remnant are used to evaluate an X-ray spectrum.

The problem here can arise in the correct hydro calculations without knowing
the ionization stage of gas.
For CSM, if it consists mostly of hydrogen, ionization should not affect
hydro evolution strongly.
But the ejecta is more metal-abundant.
Number of free electrons can vary by a factor of a few for different
ionization stages, that leads to strong changes in pressure within the
ejecta and, therefore, to possible differences in the speed of the reverse
shock.
Because of limited computational resources we decide to start with flow models
with a simplified equation of state. We assume a fixed, but not very low or
very high, ionization stage (say, the 8th) of shocked gas
during the calculations.
In future we plan to iterate the ionization stage and repeat hydro
calculations, which should make them more self-consistent.

\section*{Results and comparison with observations of Tycho remnant}

The resulting hydrodynamical structure of our models at the age of Tycho
remnant (430 years) is shown in \rfig{p_mr} and \rfig{dd_w}.
In each model the reverse shock is formed and goes inward the ejecta,
but does not reach the center yet.

\FIGG p_mr {p800_5d24_tdv} {mr_800_5d24_tdv} [width=.5\textwidth]
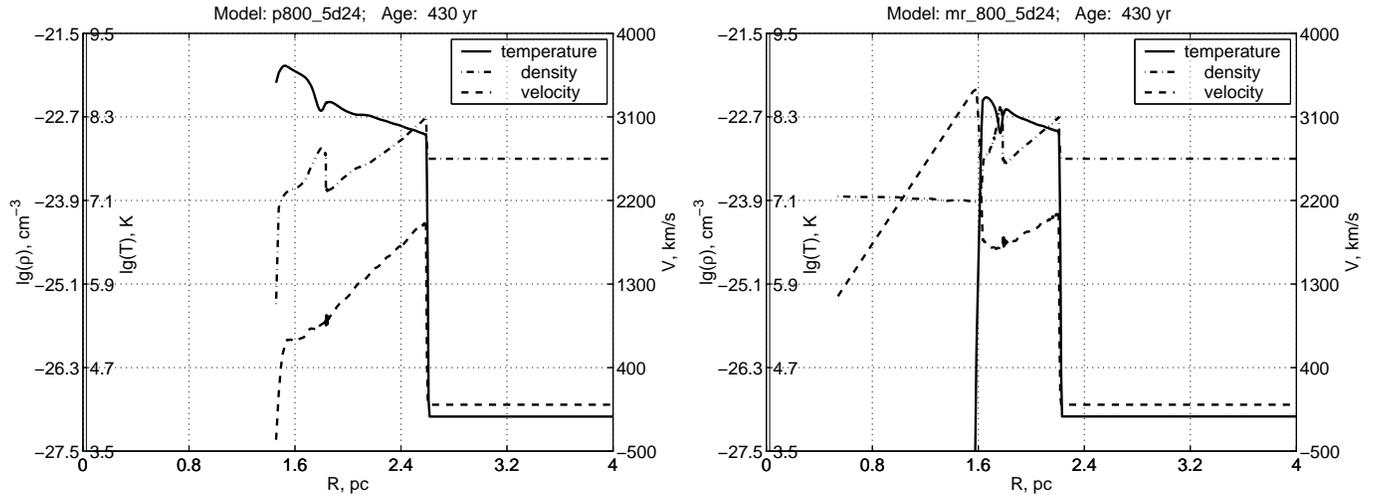
Temperature (solid), density (dash-dotted) and velocity (dashed) profiles
for models: {\bf left:} WD065; {\bf right:} MR0

\FIGG dd_w {dd4_800_5d24_tdv} {w7_800_5d24_tdv} [width=.5\textwidth]
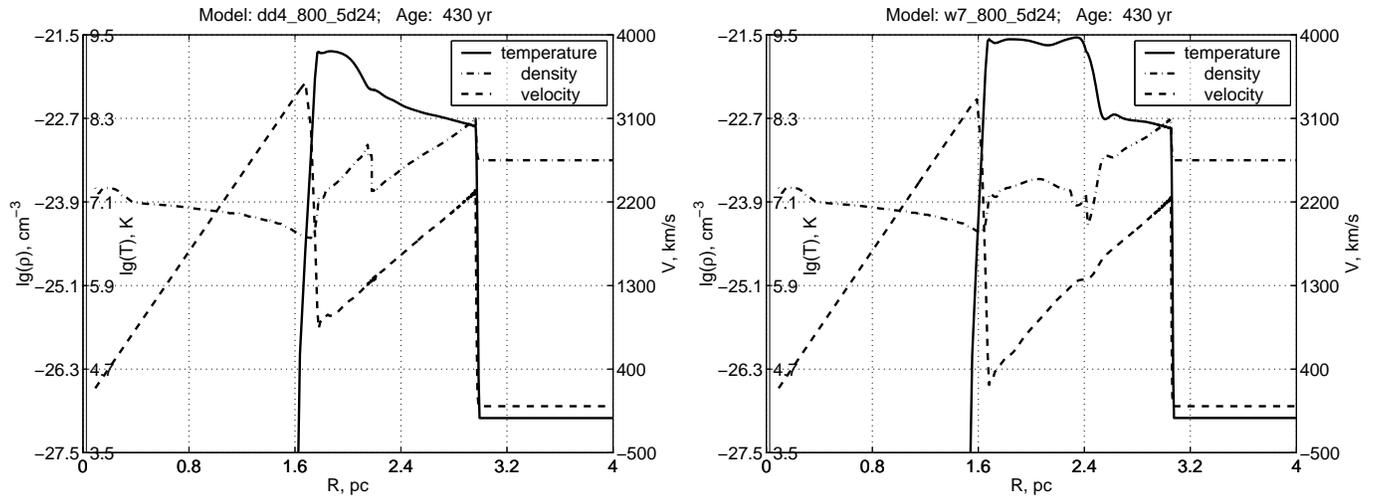
Temperature (solid), density (dash-dotted) and velocity (dashed) profiles
for models: {\bf left:} DD4; {\bf right:} W7

Due to different explosion energy in our models, they expand to the given age
till different radii, the position of the contact discontinuity
(that separates ejecta and ambient medium) for different explosion
models also varies from 1.8 to 2.5~pc. Maximum value of the shocked
ejecta density can vary in a range of an order of magnitude for
different models. The differences in temperature of heated ejecta are of the
order of half magnitude for different models. The width of the heated
(therefore, emitting) region varies also quite noticeably. All these
varieties lead to the differences in the emission one could observe from the
remnant.

Results of the spectra simulations for different explosion models
are shown in \rfig{p_mr_sp} and \rfig{dd_w_sp}. The gray line
corresponds to $N_H=4\times10^{21}\t{cm}^{-2}$ galaxy column
density, the black one -- to $N_H=9\times10^{21}\t{cm}^{-2}$.

These plots show that we can rule out the WD065 model (it shows
very weak Fe K--line, contrary to the EPIC observation with XMM--Newton by
Decourchelle et al. 2001). All Chandrasekhar-mass models produce almost the
same spectra with similar strength of lines. Though Fe K emission is a bit
stronger in MR0, it is not possible to judge which of these models is
better.

\FIGG p_mr_sp {p800_5d24abs_430_fsp} {mr_800_5d24abs_430_fsp}
[width=.5\textwidth] Full theoretical X-ray spectra (with the
account of interstellar absorption for column density $N_H$) of
the simulated remnant in XMM-Newton range (0.2-10 keV) for various
types of models: {\bf left:} WD065; {\bf right:} MR0

\FIGG dd_w_sp {dd4_800_5d24abs_430_fsp} {w7_800_5d24abs_430_fsp}
[width=.5\textwidth] Full theoretical X-ray spectra (with the
account of interstellar absorption for column density $N_H$) of
the simulated remnant in XMM-Newton range (0.2-10 keV) for various
types of models: {\bf left:} DD4; {\bf right:} W7.

Much more drastic differences arise when we look at the
profiles of the remnant in bandwidths of various
ion lines and compare these with observed azimuthally averaged
radial profiles of the deconvolved images in various lines
(Decourchelle et al. 2001).

We plot (\rfig{p_mr_pr1} and \rfig{dd_w_pr1}) the profiles of the
simulated remnant in different lines (e.g., in FeXVII line and Si~K
line). Models DD4 and W7 give comparable profiles, but Si and Fe lines are
separated strongly in space, WD065 is
distinguished strongly from the others and from the observational
data, while MR0 shows the best agreement with observations: the diameters of
the remnant in Si and Fe lines are very close,
an amplitude of the luminosity variations in the corresponding lines is also
more similar to what is observed by Decourchelle et al. 2001.

\FIGG p_mr_pr1 {p800_5d24_2_3_3_430_xmm}
{mr_800_5d24_2_3_3_430_xmm} [width=.5\textwidth] Profiles of the
remnant's brightness in lines of Si~K  and  FeXVII ions,
integrated over line of sight. Dashed line shows a profile of
emission within a range 1.67-2.0 keV (Si~K) and 775-855 eV
(FeXVII). Solid line shows emission within a range of one bin
($\Delta E = 13.6$ eV @ 1 keV) that presumably corresponds to the
emission line of Si~K or FeXVII. {\bf Left:} WD065; {\bf right:}
MR0

\FIGG dd_w_pr1 {dd4_800_5d24_2_3_3_430_xmm}
{w7_800_5d24_2_3_3_430_xmm} [width=.5\textwidth] Profiles of the
remnant's brightness in lines of Si~K  and  FeXVII ions,
integrated over line of sight. Dashed line shows a profile of
emission within a range 1.67-2.0 keV (Si~K) and 775-855 eV
(FeXVII). Solid line shows emission within a range of one bin
($\Delta E = 13.6$ eV @ 1 keV) that presumably corresponds to the
emission line of Si~K or FeXVII. {\bf left:} DD4; {\bf right:} W7

\section*{Discussion and prospects}
Though the results are very preliminary, we can rule out the
sub-Chandrasekhar-mass model WD065
for the Tycho SNR. The most popular models DD4 and W7 show
relatively good agreement with observational data (though the
radial profiles are not very encouraging). Nevertheless, the most
promising model is still MR0. It fits better both the observed X-ray
spectrum and the brightness profiles in different lines.

To make our results more reliable, we still need to
upgrade our techniques in order to obtain a better agreement with X-ray
observations. There is a number of issues that have to be worked out.

Since we do not know exactly the ionization stage (thus the
``true'' equation of state) of plasma on the hydro calculation
step, we should calculate it by our time-dependent code. But due
to our limited computer resources we cannot make it in a fully
self-consistent way immediately. Instead, we can calculate
separately and save the history of ionization and then perform the
second iteration of flow calculation with the new (updated) set of
ionization data.

In the modelling we assume $T_e=T_i$ in the calculations. We can
introduce a parameter that specifies relationship between $T_e$
and $T_i$. Variations of this parameter will affect the spectrum
shape. Thus we can find a value of the parameter that will make a
best fit to spectrum.

We have only tested CSM with a constant density. In near future we plan to
test also ``windy'' CSM, with density gradient. That will change the
dynamical evolution and, probably, the emission of our models.

At last, looking at the spectra one can see, that Decourchelle et
al. 2001 show very broad lines and we have them rather narrow,
although we have taken into account thermal and Doppler
broadening. Convolving our theoretical curves with XMM-Newton
response matrix (EPIC, MOS and PN instruments) would give us a
possibility for more accurate data fitting.

Taking into account all the above issues we will
make the comparison of models and observations more correct and
fruitful. Nevertheless, we believe that the main conclusion of this work
will remain unchanged: the new 3D SN~Ia model MR0, due to its low energetics
and a strong mixing, seems the best to model Sn~Ia remnants, as well as
their light curves.

{\it Acknowledgements.} The work is supported in Russia by RFBR grant
02-02-16500.

\vskip 2mm
E-mail address of D.I.~Kosenko lisett@xray.sai.msu.ru\\
Manuscript received \phantom{19 October 2002};
revised \phantom{19 October 2002};
accepted \phantom{19 October 2002}.


\begin{thebibliography}{}
\let\bibitem=\item
\bibitem Baade~W., {\it Astroph. J.}, {\bf 102}, 309, 1945.
\bibitem Blinnikov,~S.I., Bartunov,~O.S., {\it Astron. Astrophys.},
{\bf 273}, 106, 1993.
\bibitem Blinnikov,~S.I., Sorokina~E.I., Hunting the Cosmologocal
Parameters, Eds. D.~Barbossa et al., {\it KLUWER}, 2003.
\bibitem Blinnikov,~S.I. et al.; {\it Astroph. J.}, {\bf 532}, 1132,
2000.
\bibitem  Borkowski,~K.J., Lyerly,~W.J., and Reynolds,~S.P.,
{\it Astroph. J.}, {\bf 548}, 820, 2001.
\bibitem Brinkmann,~W., Fink,~H.H., Smith,~A., and  Haberl,~F.,
{\it Astron. Astrophys.}, {\bf 221}, 385, 1989.
\bibitem Chevalier~R.A., Raymond~J.C., {\it Astroph. J. Lett.}, {\bf 225},
L27, 1978.
\bibitem Davison~P.J.N., Culhane~J.L., Mitchell~R.J., {\it Astroph. J. Lett.},
{\bf 206}, L37, 1976.
\bibitem Decourchelle~A. et al.,{\it Astron. Astrophys.}, {\bf 365}, L218,
2001.
\bibitem Dickel~J.R. et al., {\it Astroph. J.}, {\bf 257}, 145, 1982.
\bibitem Duin~R.M., Strom~R.G., {\it Astron. Astrophys.}, {\bf 39}, 33,
1975.
\bibitem Hamilton~A.J.S., Sarazin~C.L., Szymkowiak~A.E., {\it Astroph. J.},
{\bf 300}, 713, 1986.
\bibitem Hwang~U., Gotthelf~E.V., {\it Astroph. J.}, {\bf 475}, 665, 1997.
\bibitem Hwang~U., Hughes~J.P., Petre~R., {\it Astroph. J.}, {\bf 497}, 833,
1998.
\bibitem Itoh~H., Masai~K., Nomoto~K., {\it Astroph. J.}, {\bf 334}, 279,
1988.
\bibitem Nomoto,~K., Thielemann,~F.K., Yokoi,~K., {\it Astroph. J.},
{\bf 286}, 644, 1984.
\bibitem Reinecke,~M., Hillebrandt,~W., Niemeyer,~J.~C.,
{\it Astron. Astrophys.}, {\bf 386}, 936, 2002.
\bibitem Ruiz-Lapuente,~R. et al., {\it Nature}, {\bf 365}, 728,
1993.
\bibitem Seward~F., Gorenstein~P., Tucker~W., {\it Astroph. J.}, {\bf 266},
287, 1983.
\bibitem Tsunemi~H et al., {\it Astroph. J.}, {\bf 306}, 248, 1986.
\bibitem van den Bergh~S., {\it Astroph. J.}, {\bf 168}, 37, 1971.
\bibitem van den Bergh~S., Marscher~A.P., Terzian~Y., {\it Astroph. J.
Suppl.}, {\bf 26}, 19, 1973.
\bibitem Vancura~O., Gorenstein~P., Hughes~J.P., {\it Astroph. J.},
{\bf 441}, 680, 1995.
\bibitem Woosley,~S.E., Weaver,~T.A., Supernovae, Eds.
J.~Audouze et al., {\it ELSEVIER}, Sci.Pub., Amsterdam, 63, 1994.

\end{thebibliography}
\end{document}